\documentclass[twocolumn,10pt]{IEEEtran}
\usepackage{indentfirst}
\usepackage{amsmath}
\usepackage{amsfonts}
\usepackage{amssymb}
\usepackage{cite}
\usepackage{times}
\usepackage[dvips]{graphicx}
\usepackage{hhline}
\usepackage{color}
\usepackage{multirow}
\usepackage{graphicx}
\usepackage{epstopdf}
\usepackage[top=0.7in,bottom=0.95in, left=1.8cm, right=1.8cm]{geometry}
\usepackage{soul}

\newcommand{\bi}{\begin{itemize}}
\newcommand{\ei}{\end{itemize}}

\newcommand{\be}{\begin{IEEEeqnarray}}
\newcommand{\ee}{\end{IEEEeqnarray}}

\newcommand{\commentLB}[1]{}

\usepackage[absolute]{textpos}
\usepackage{everyshi}

\begin{document}
	
	\title{Information Theoretic Key Agreement Protocol based on ECG signals}

	\author{ 
		Anna V. Guglielmi$^{1,2}$,\thanks{This work was supported by GARR consortium. Part of this work was also supported by MUR (Ministry of University and Research) under the initiative Departments of Excellence (Law 232/2016).}
		Alberto Muraro$^{2}$,
		Giulia Cisotto$^{2,3,4}$,
		\and and Nicola Laurenti$^{2,4}$ \\
		$^{1}$Gestione Ampliamento Rete Ricerca (GARR) Consortium, Rome, Italy\\
		$^{2}$Dept.\ of Information Engineering, University of Padova, Italy \\
		$^3$National Centre for Neurology and Psychiatry, Tokyo, Japan \\
$^4$National Inter-University Consortium for Telecommunications (CNIT), Padova, Italy\\
		email: \{guglielm, muraroalbe, nil\}@dei.unipd.it, giulia.cisotto.1@unipd.it
	}

	\date{}
	\maketitle
	\thispagestyle{empty}
	\pagestyle{empty}

\begin{abstract}
Wireless body area networks (WBANs) are  becoming increasingly popular as they allow individuals to continuously monitor their vitals and physiological parameters remotely from the hospital. With the spread of the SARS-CoV-2 pandemic, the availability of portable pulse-oximeters and wearable heart rate detectors has boomed in the market. At the same time, in 2020 we assisted to an unprecedented increase of healthcare breaches, revealing the extreme vulnerability of the current generation of WBANs. Therefore, the development of new security protocols to ensure data protection, authentication, integrity and privacy within WBANs are highly needed.
Here, we  targeted  a  WBAN  collecting  ECG  signals from different sensor nodes on the individual's body, we extracted the inter-pulse interval (i.e., R-R interval) sequence from each of them, and we developed a  new  information  theoretic  key  agreement  protocol that exploits the inherent randomness of ECG to ensure authentication between sensor pairs within the WBAN.
After proper pre-processing, we provide an analytical solution that ensures robust authentication; we provide a unique information reconciliation matrix, which gives good performance for all ECG sensor pairs; and we can show that a relationship between information reconciliation and privacy amplification matrices can be found. 
Finally, we show the trade-off between the level of security, in terms of key generation rate, and the complexity of the error correction scheme implemented in the system.
\end{abstract}

\begin{textblock*}{17cm}(1.7cm, 0.5cm)
\noindent\scriptsize This work has been submitted to the IEEE for possible publication. Copyright may be transferred without notice, after which this version may no longer be accessible.\\
\textbf{Copyright Notice}: \textcopyright 2021 IEEE. Personal use of this material is permitted. Permission from IEEE must be obtained for all other uses, in any current or future media, including reprinting/republishing this material for advertising or promotional purposes, creating new collective works, for resale or redistribution to servers or lists, or reuse of any copyrighted component of this work in other works.
\end{textblock*}

%
%
%
%


\section{Introduction}
\label{sec:sec0}

\IEEEPARstart{W}{ireless} Body Area Networks (WBANs) are used for communication between intelligent, miniaturized, low-power biosensors that operate in and/or on the human body in order to gather patients’ medical information (e.g., heart rate via electrocardiogram (ECG), brain activity via electroencephalogram (EEG), oxygen saturation level via photoplethysmographic (PPG), body movement, body temperature, blood pressure, etc.) and eventually generate alarms in case of health hazards \cite{Movassaghi2014, Martiradonna2021}.
WBANs represent only the first tier in a typical e-health system, including also the inter-WBAN tier (e.g., aggregating data from different WBANs), and the beyond-WBAN tier (e.g., cloud storage) \cite{Movassaghi2014, AESM2018, Martiradonna2021}. Then, each tier can be affected by specific security issues, as systematically reviewed by \cite{Hathaliya2020}.
As WBANs become increasingly popular, and the number of biosensors and wearables on the market is constantly growing, the security aspects related to, e.g., data privacy, user authentication, and data integrity are emerging as critical issues to cope with since the WBAN design phase \cite{Aljanabi2017,Hathaliya2020}. 
Furthermore, it has been widely recognised that SARS-CoV-2 pandemic has led to a significant increase of phishing campaigns, mishandled health record disposals, and sophisticated cyber-attacks \cite{Wowmom2016} \cite{cyberattack2020}: in fact, they represent some of the biggest healthcare data breaches of 2020. It has been estimated that 41.4 million patient records has been breached in 2019, with a 49$\%$ increase in hacking (according to the Protenus Breach Barometer). Such evidences have strongly highlighted the security vulnerabilities of the healthcare sector and its information systems.
%
%
The data collected via WBAN are sensitive and can be subjected to several well-known security issues, e.g., man-in-the-middle attacks, cross-site request forgery, spoofing, or pishing \cite{chaudhary}. Besides, wearables and portable devices to measure biometric parameters have stringent resource limitations, concerning energy, transmission and storage. Therefore, security solutions should be lightweight as well as efficient.



In this paper, we target a WBAN collecting ECG signals from different locations on the individual's body, and we propose a new information theoretic key agreement protocol that exploits the inherent randomness of ECG signals to ensure authentication between the nodes within the WBAN.


Compared to prior works, we assume no side information model is available, and we show that we can generate a secure and secret keys by properly applying information reconciliation and privacy amplification techniques, exploiting error correction strategies, the design of proper parity-check and privacy amplification matrices \cite{maurer}, \cite{bennet}. 
As in previous literature, from the ECG signals, we extracted the cardiac inter-pulse interval (IPI), i.e., the time difference between two consecutive heart beats.
IPI can be considered as a time- and person-specific parameter and, as such, it allows the development of authentication and, more generally, security protocols exploiting its inherent randomness over time.
Each IPI is transformed into a sequence of random bits, and we assume that legitimate sensors, only, can retrieve that sequence, by simultaneously measuring ECG from two different locations and applying the proposed key agreement protocol.
Previous literature has already proposed IPI-based security mechanisms: they mainly employed fuzzy commitment and fuzzy vault \cite{seepers, cho, choi, kim}.
%
%
On the contrary, we provide an analytical solution to authenticate two nodes within a WBAN: we can show that a relationship between information reconciliation and privacy amplification matrices can be found.
Furthermore, we can design a unique information reconciliation matrix, which provides good performance for all the considered ECG sensors. In fact, an ECG sensor-specific information reconciliation matrix can be defined such that to exploit the correlation between different pairs of ECG sensors.

The rest of the paper is organized as follows. In Section \ref{sec:relwor}, prior work on security protocols for WBANs is discussed. Section \ref{sec:sysmod} describes the system model, including the scenario, the IPI extraction, and the key agreemeent protocol development. Numerical results are presented and discussed in Section \ref{sec:numres}. Finally, Section \ref{sec:conc} concludes the paper.

\section{Related works}
\label{sec:relwor}

The rapid penetration of wearable and portable devices to monitor physiological parameters and vitals is posing an increasing need for security protocols for WBAN. Very recent and exhaustive surveys reported detailed discussions on authentication schemes, security protocol design, and development issues in WBANs \cite{joshi}, \cite{chaudhary}, and \cite{hussain}.
Currently, one of the most common solutions to provide security in WBAN is based on the identification of unique biometric traits, e.g., voice, fingerprint, iris \cite{hussain, Aljanabi2017} and their use to generate cryptographic keys. The authentication of the user of the WBAN allows to securing communications between the sensing device (i.e., single-lead ECG, pulse-oximeter) and the gateway device (e.g., the smartphone or the personal digital assistant), as well as to ensure that data transmitted to the user's caregiver (located in a different place, beyond the WBAN) are secure \cite{Aljanabi2017}.

However, the abovementioned solutions make use of stationary traits, i.e., the iris or the fingerprints can have a limited range of variation across days or conditions: this ensures robustness to the system; but, at the same time, it can make the system prone to possible attacks if a malicious individual obtains such information.
Therefore, recent works have proposed the use of dynamic biometric traits, i.e., physiological signals, such as ECG \cite{hu, zhou} and PPG \cite{ven} to generate cryptographic keys. Not only on-body WBANs, but also in-body WBANs could benefit of such recent approaches.
%
For instance, in \cite{H2H}, ECG is used for the authentication to an IMD called Heart-to-Heart (H2H). The security scheme needs a touchable medical instrument that must be in physical contact with the user, so that it can accept and authorize access to the user to read the signals acquired by the IMD. The described mechanism compares the user’s ECG from the touchable medical instrument and the IMD in order to access the IMD data.
Similarly, in \cite{conti}, ECG is employed, together with EEG, for the authentication to another IMD.

Dynamic biometric traits are characterized by a higher level of inherent randomness, with respect to a stationary one: thus, this can be exploited to provide additional security within the WBAN.
However, at the same time, this implies further complexity in the processing of such signals. Therefore, often, expert features are extracted from the signal, prior to implement the security protocol.
%
As an example, it is common to use the IPI, extracted from an ECG signal, to provide authentication in a WBAN.

%
In \cite{IMDGuard}, the authors introduced a security scheme for IMDs with comprehensive techniques for the ECG-based keys using IPI with secure protocol and on the access control mechanism on the IMD external devices with an authentication proxy to protect it. 
Recently, \cite{seepers} presented a new IPI-based key-exchange protocol employing fuzzy commitment to tolerate the expected disparity between IPIs obtained by an external reader and an IMD, as well as a novel way of tackling heart beat mis-detection through IPI classification. They found that roughly one minute is the expected time for securely exchanging an 80-bit key with high probability (1-10$^{-6}$), while consuming only 88 $\mu$J from an IMD. Some improvements (both in terms of computational and communication cost and key establishment time) of the abovementioned scheme were reported in \cite{cho, choi, kim}.
In particular, in \cite{cho}, fuzzy-vault based protocol was shown to shorten the key establishment time by reducing the number of IPIs needed to generate a secret key. 
In \cite{choi}, an energy-efficient key exchange
protocol using IPIs was able to generate a secret key in an authenticated and transparent manner, without any key material being exposed
before distribution or during initialization. The authors could dramatically reduce energy consumption (in communication and computation) compared to the state-of-the-art (e.g., \cite{seepers}), while keeping secure key exchange.
Finally, in \cite{kim} the authors proposed an improvement over the approach described in \cite{seepers}, by introducing a new method to detect mis-detections and recover IPIs using thresholds. They could show a significant reduction of the cost for communications (which was not only energy efficient, but it was able to generate secure keys). Their results showed that the total number of IPI recoveries is 19, while the key generation rate is 99.4$\%$.

\section{ECG-based key agreement protocol}
\label{sec:sysmod}

\subsection{Scenario}
\label{subsec:scenario}
\begin{figure}[h!]
\centering
\includegraphics[scale=0.16]{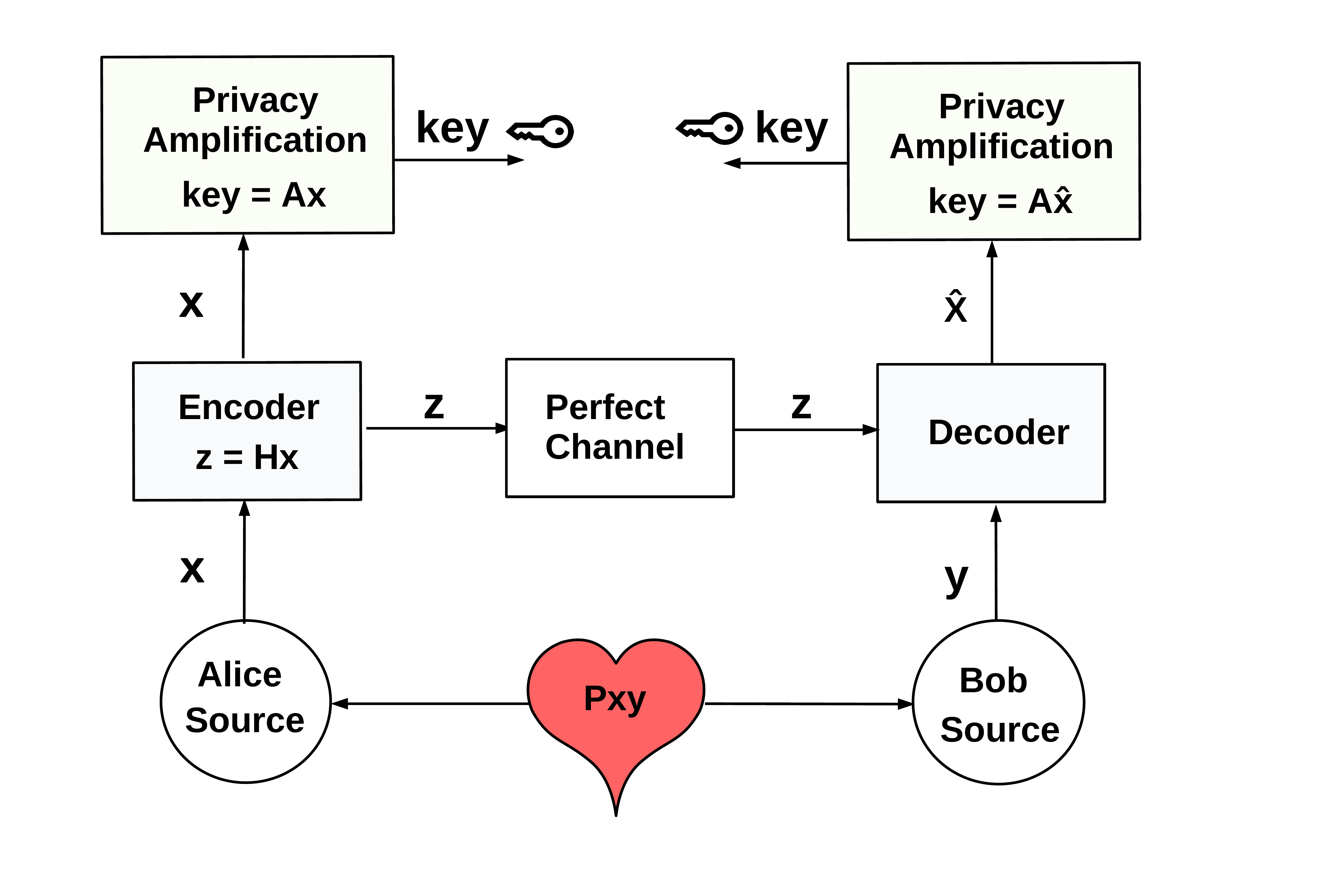}
\caption{Key agreement scenario with correlated sources for Alice and Bob. Their correlation is expressed by the joint PDF $p_{xy}$.}
\label{fig:inforecscenario}
\end{figure}

Fig. \ref{fig:inforecscenario} shows the scenario taken into account. We consider two legitimate sensors, named Alice and Bob, belonging to the on-body network and collecting the authentic individual’s ECG signals, indicated as $x$ and $y$, respectively.
We assume that an eavesdropper, i.e., Eve, is a potential attacker that cannot observed  $x$ or $y$, nor any other ECG signal correlated with them, but can read the information exchanged by Alice and Bob on the noiseless and authentic but otherwise insecure public channel.

We aim at designing a robust key agreement security protocol exploiting the inherent correlated randomness of ECG signals to ensure authentication, confidentiality, integrity, and accountability within the on-body sensors network. Indeed, the proposed key agreement protocol involves the joint generation of a secret key by Alice and Bob based on $x$ and $y$, through the exchange of information over the public channel.

When dealing with the design of such protocol, the first step is to evaluate the \emph{secret key capacity} for the considered scenario, in order to define the maximum rate at which Alice and Bob can generate the secret key by communication over the public channel. Our scenario can be cast into what is called the ``source-type model with no side information for the attacker'' in \cite{AhCs1993}, and therefore the secret key capacity is given by the \emph{mutual information rate} (MIR)
\begin{equation}
    C_{\rm k} = \frac1T I(x;y)
\end{equation} 

In the practical three-step implementation outlined in \cite{bennet} secret key agreement, Alice and Bob then have to reconcile their random but correlated information. In other words, they have to correct the discrepancies in their observations by exchanging additional error correction information as discussed in \ref{subsec:ir}. Then, they have to distill secret bits from the corrected data using a technique named privacy amplification, as described in \ref{subsec:pa} \cite{bennet}, \cite{maurer}. 

In our preliminary observations, we found that the ECG can be hardly characterized by using well-known statistical distributions. Previous literature provided tools (based on Matlab) and theoretical approaches that lead to very long execution times, high computational complexity, and inaccurate MIR estimate. Thus, we considered an empirical (i.e., data-driven) modeling of such signals obtaining that the estimate of MIR upper and lower bounds can be derived. In particular, the MIR between two Gaussian processes that have the same second order joint statistics of the considered ECG represents a lower bound of the true MIR. On the other hand,  the single symbol pairs mutual information between the two ECG is an upper bound to their MIR. 

\subsection{ECG signals and IPIs}
\label{subsec:data}

\begin{figure}[h!]
\centering
\includegraphics[scale=0.8]{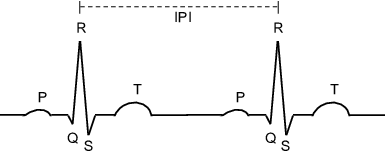}
\caption{Representation of two consecutive ECG segments and their IPI evaluation.}
\label{fig:ipi}
\end{figure}

ECG measures the heart activity by acquiring the voltage difference between two locations over the individual's body. The ECG electrodes detect the small electrical changes that are a consequence of cardiac muscle depolarization followed by repolarization during each cardiac cycle, i.e., the heart beat. The resulting ECG is a quasi-period signals that includes a well-known sequence of peaks, \emph{waves} and \emph{segments} in each of its periods. Fig. \ref{fig:ipi} represents a schematic ECG signal with two consecutive heart beats. Among other components, the R peak identifies the heart beat and it is commonly used to compute the heart rate, i.e., as the time difference between two consecutive R peaks.

In order to extract the IPI feature (used as input to our key agreement protocol), first the R peaks are identified along the entire signal (the same pre-processing is performed at Alice's and Bob's sensors node).
Given the sequence of R peaks, each IPI value is obtained by calculating the time difference between two consecutive R peaks, as shown in Fig. \ref{fig:ipi}.



For the purpose of our work, we define as $x$ and $y$ the sequences of IPIs extracted by the two legitimate sensor nodes within the WBAN (i.e., instead of the raw ECG signals).


\subsection{Quantization} 
\label{subsec:quant}
A necessary step before processing is quantization. When dealing with continuous signals such as physiological signals, they need to be analog-to-digital converted (ADC) to obtain discrete values (symbols) and generate the secret key. For the considered ECG signals the native sensor quantization uses $b=16$ bits/sample, while the IPI is inherently quantized with a step given by the ECG sampling period $T_s$. 

For good quality sensors with high sampling rate, the time resolution of IPIs is too fine and yield frequent different values between $x$ and $y$. Therefore, we aim at a coarser quantization that, without sacrificing the key rate, reduces the discrepancy occurrences.  

Our aim is therefore to design a quantizer which uses few bits to represent each IPI and at the same time does not significantly decrease $I(x,y)$, the mutual information between $x$ and $y$. 

Let $Q_x$ and $Q_y$ be Alice and Bob's quantizers and $\widetilde{x} =  Q_x(x)$ and $\widetilde{y} =  Q_y(y)$ the quantized version of $x$ and $y$, with $\widetilde{x},\widetilde{y} \in \{1,2,\dots ,L\}$.
Let us define $x^\ast$ as the coincidence values between $\widetilde{x}$ and $\widetilde{y}$
\begin{equation}
  x^\ast = 
  \begin{cases}
    \widetilde{x} & \text{if }\widetilde{x}=\widetilde{y} \\
    \emptyset & \text{otherwise}
  \end{cases}
\end{equation}
Then the probability mass distribution (PMD) of $x^{\ast}$ is
\begin{equation}
p_{x^\ast}(\ell) = \int_{R_\ell^{x}\times R_\ell^{y}} p_{xy}(a,b)\,da\,db ,\quad \ell \in \{1, \ldots, L\}
\end{equation}
where $R_\ell^{x}$ and $R_\ell^{y}$ are Alice and Bob's quantizer regions associated to $\ell\in\{1,\dots ,L\}$ and $P\left[ x^\ast = \emptyset \right] = 1-\sum_\ell p_{x^\ast}(\ell)$.
Our aim is to maximize the entropy times frequency of $x^\ast$
\begin{equation}
P\left[ x^\ast \neq \emptyset \right] H(x^\ast) = \sum_{\ell=1}^{L} p_{x^\ast}(\ell) \cdot \log_{\frac{1}{2}}\left(\frac{p_{x^\ast}(\ell)}{P\left[ x^\ast \neq \emptyset \right]} \right).
\end{equation}
The optimal quantizer threshold values can thus be found by numerical optimization from the joint PDF. An efficient recursive implementation would reuse the optimal $\ell$-th threshold for $L$ levels, $\tau_{\ell, L}$ as the optimal $\tau_{2\ell, 2L}$ and optimize each $\tau_{2\ell+1, 2L}$ threshold between $\tau_{2\ell, 2L}$ and $\tau_{2\ell+2, 2L}$.
Moreover, since an analytical model for $p_{xy}$ is not available we make use of the empirical distribution derived from data, replacing $p_{x^\ast}(\ell)$ with $N_\ell/N$ where $N_\ell=|\{(x_i,y_i) \in R_l^{x}\times R_l^{y}\}|$ 

\begin{figure}[h!]
\centering
\includegraphics[scale=0.45]{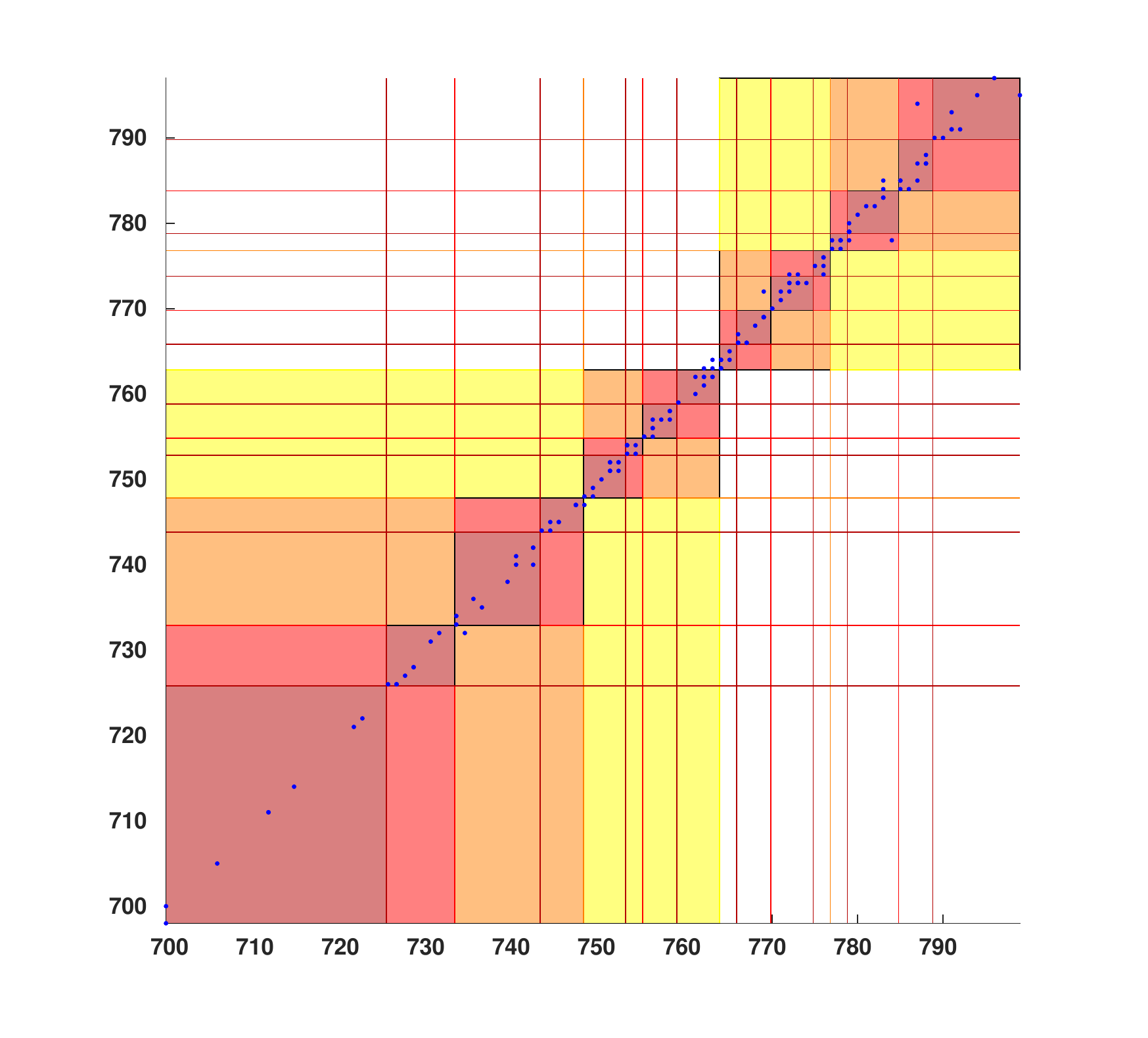}
\caption{IPI quantization thresholds for 1- (yellow), 2- (orange), 3- (red) and 4-bit (dark red) quantization. Values are in ms}
\label{fig:joint_4bit}
\end{figure}

Fig. \ref{fig:joint_4bit} shows the different stages of the recursive procedure, where the colors get darker as the number of quantization bits increases. For example the two yellow quadrants represent the quantization with 1 bit where there is only a delimiter for each variable.


We must observe that, given the high correlation of IPI sequences, even simplfying the choice to the non uniform quantizers that yield uniformly distributed values $p_{\tilde x}(\ell) = p_{\tilde y}(\ell) = 1/L$ yields results that are only slightly worse (not shown here).

\subsection{Information reconciliation}
\label{subsec:ir}
During information reconciliation, 
Alice compresses its source symbols $\tilde{x}$ indicated as $z$ in Fig. \ref{fig:inforecscenario} and transmits them over the public channel.  
Bob decodes $z$ with the help of correlated side information, that is $\tilde{y}$.  
The goal is to allow Bob to correct the discrepancies between his symbols $\tilde{y}$ and $\tilde{x}$ minimizing $z$, that is those bits that can be seen by Eve. In this way, Alice and Bob should be able to independently generate the same secret key based on $x$ and the corrected version of $y$.

 

An Hamming distance-based syndrome decoding scheme has been considered to do error correction. Defined a parity-check matrix $\mathbf{H}$ of size NxM, with N$>$M, the syndrome $z$ of length M is computed as $\mathbf{H} \cdot \tilde{x}$. According to the received $z$, Bob builds the set $\mathcal{W}$ of all the possible words with $z$ as syndrome
\begin{equation}
\mathcal{W} = \{w \mid \mathbf{H}\cdot w= z\}
\end{equation}
choosing then the one, i.e., $w$, with the minimum Hamming distance from $\tilde{y}$. 

To evaluate the information reconciliation performance, we compute the percentage of different bits per symbol between Alice symbols and Bob decoded symbols.


\subsection{Privacy amplification}
\label{subsec:pa}
Once the discrepancies between Alice and Bob's observations have been corrected by means of information reconciliation, the secret bits have to be distilled from the corrected data. Note that Eve only knows $z$ since it is transmitted over the public channel, that is the M bits transmitted by Alice to Bob. 
For this reason, the M bits must be removed from $\tilde{x}$ and $\hat{x}$ before the generation of the secret key. 
To do so, a privacy amplification matrix $\mathbf{A}$ must be designed based on the knowledge of $\mathbf{H}$ so that $key=\mathbf{A}\tilde{x}^{T}=\mathbf{A}\hat{x}^{T}$, where $\tilde{x}$ is Alice's quantized message, $\hat{x}$ is Bob's message after the reconciliation and the $key$ has length K. It is worth noting that the key's bits should follow a uniform distribution 
(i.e., to assure perfect key bits uniformity) and $I(key,z)=0$, that is the mutual information between the generated key and $z$ should be zero (i.e., to assure perfect secrecy).

$\mathbf{A}$ represents an injective map: many keys are related to the same syndrome. Moreover, $\mathbf{A}$ should be such that the key set is uniformly distributed given Eve's knowledge. In mathematical terms, this means that the null space of $\mathbf{A}$, denoted as $\mathcal{N}(\mathbf{A})$, must be orthogonal to the null space of $\mathbf{H}$, i.e., $\mathcal{N}(\mathbf{H})$. In particular, the elements of $\mathcal{N}(\mathbf{H})$ must be uniformly mapped into the set $\{0,1\}^K$. The elements of $\mathcal{N}(\mathbf{H})$ are in $\{0,1\}^N$.


Assuming that $\mathbf{\mathbf{H}}$ is full rank, then rank($\mathbf{H}$) = M and from the rank–nullity theorem
\begin{equation}
    rank(\mathbf{H}) + \mathcal{N}(\mathbf{H}) = N
\end{equation}
meaning that $\mathcal{N}(\mathbf{H}) = K$. In other words, $2^K$ vectors in $\{0,1\}^N$ need to be mapped into $2^K$ vectors in $\{0,1\}^K$. 
This function must be bijective assuring the uniformity of the generated key.

Finally, if we define $r=\frac{N-M}{N}$ as the rate of the considered code, $b$ the quantization bits and $T$ the average time (in seconds) between two consecutive R peaks, we have that the rate of key generation is  $r\cdot b \cdot \frac{1}{T}$ [bit/s].


\section{Results and discussion}
\label{sec:numres} 

The dataset used is the PTB ECG diagnostic dataset accessible online at \cite{ecg}. It consists of $549$ records from $290$ subjects (aged $17$ to $87$, mean $57.2$; $209$ men, mean age $55.5$, and $81$ women, mean age $61.6$). Each subject is represented by one to five records and each record includes $15$ simultaneously measured ECG signals: the conventional $12$ leads (I, II, III, AVR, AVL, AVF, V1, V2, V3, V4, V5, V6) and the $3$ Frank leads (Vx, Vy, Vz). The sampling frequency is $f_s = 1 kHz$, with $16$ bit resolution over a range of $\pm 16.384$ mV. 

The results 
have been obtained considering $4$ quantization bits and 
different $\mathbf{H}$ sizes. 
Moreover, we considered all possible combinations of paired ECG sensors,  i.e., Alice-Bob pairs. 

\begin{figure}[h!]
\centering
\includegraphics[scale=0.27]{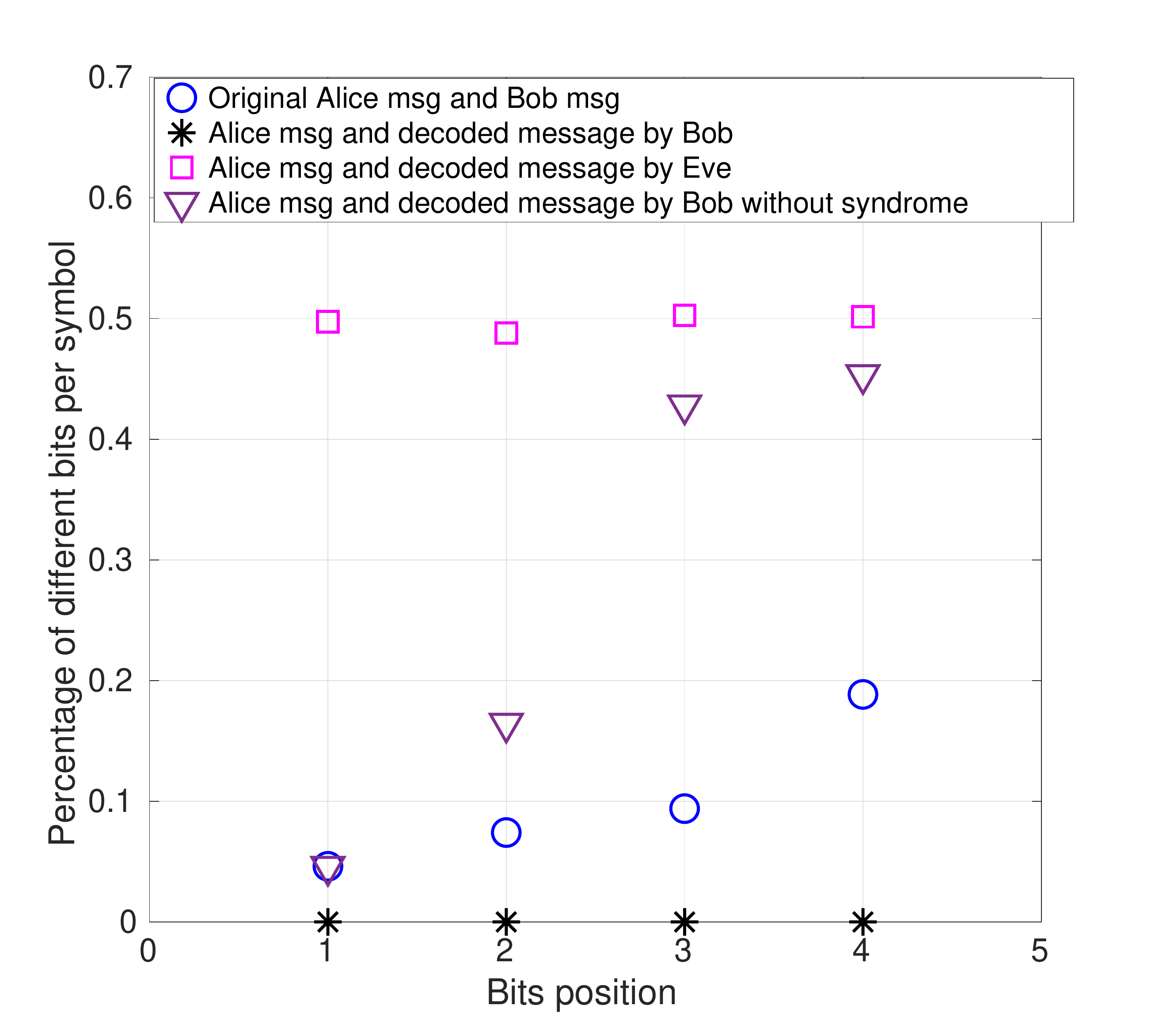}
\caption{Percentage of different bits per symbol when real ECG signals are quantized with $b=4$ quantization bits and $\mathbf{H}$ size is $160$x$142$.}
\label{fig:PTB_160x142}
\end{figure}

Fig. \ref{fig:PTB_160x142} shows the percentage of different bits per symbol when $\mathbf{H}$ size is 160x142 (averaged across all Alice-Bob pairs). It can be seen that the difference between the binary representations of Alice and Bob's original messages (i.e., symbols), right after the quantization before $z$ transmission, is never zero. As a consequence, it is reasonable to expect that Bob's decoding performance might be sub-optimal. Indeed, we compared different $\mathbf{H}$ sizes, by observing that an $\mathbf{H}$ with a higher size allows to reach a zero decoding error probability on average. 
It is also shown the difference between the binary representations of Alice's message, used for the computation of $z$, and Bob's decoded message, using $z$, and the difference between the binary representations of Alice message used to compute $z$ and Bob's decoded message using only the information provided by the joint PDF estimation based on the knowledge of the two acquired ECG signals. It can be seen that Bob takes advantage from syndrome decoding and this happens regardless of $\mathbf{H}$ size. 

In addition to this, Eve's decoding performance can be discussed as well in terms of difference between the binary representations of Alice's message and the message decoded by Eve. In this regards, it is worth underlining that the quantizer design is important not to allow Eve to empirically generate apriori knowledge about the quantized symbols and, therefore, increase its decoding performance. It can be seen that Eve's performance are around $50\%$.

In general, from the obtained results we observe that the size of $\mathbf{H}$ could be so that to have the decoding error probability as small as we want, but this could come at the expense of the total time needed to generate the final key. 
In particular, considering $\mathbf{H}$ with size 169x142, $r=\frac{160-142}{160}$, $b=4$, and $T=0.750$s, we can have on average around $0.6$ bit/s as a key generation rate.

\begin{figure}[h!]
\centering
\includegraphics[scale=0.38]{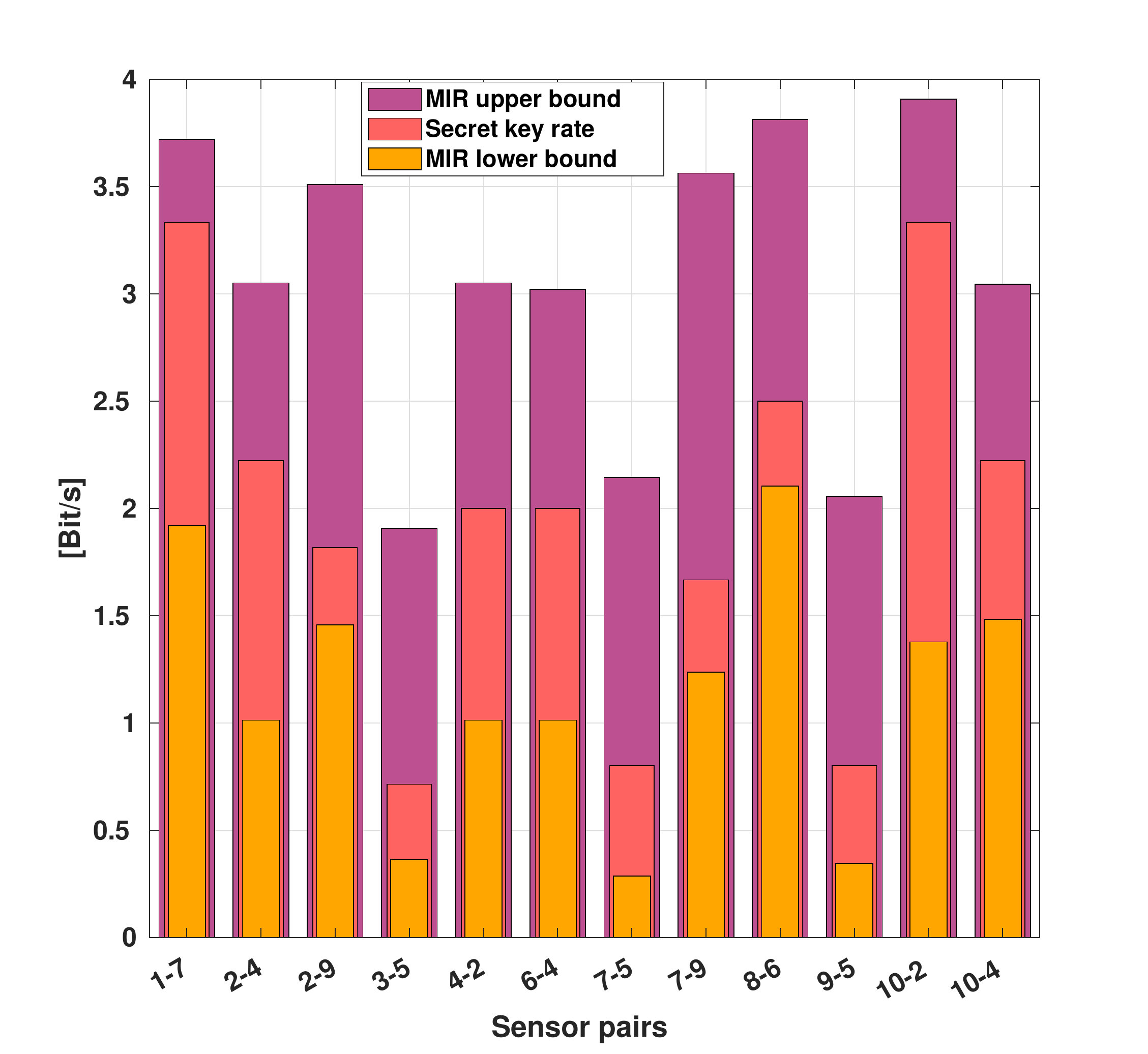}
\caption{Secret key rate with MIR upper and lower bounds. The results have been obtained varying sensors pairs and choosing the proper $\mathbf{H}$ size. }
\label{fig:MIR_Lower_upper}
\end{figure}

However, a specific $\mathbf{H}$ could be designed for each (Alice, Bob) pair in order to exploit the correlation between ECG signals acquired by different sensors. Indeed, different (Alice, Bob) couple can required a different $\mathbf{H}$ involving then a different key generation rate. 
At this regards, Fig \ref{fig:MIR_Lower_upper} shows the key generation rate [Bit/s], i.e., secret key rate, for some sensors pairs. As it can be seen, its value changes if we choose different sensors. Furthermore, it can be observed that the obtained secret key rates are in between the lower and upper bounds evaluated as discussed in Section \ref{subsec:scenario}.


In addition to the Hamming decoding approach described in Section \ref{subsec:ir}, we implemented an LDPC-based scheme as well. However, the Hamming-based approach seems to work better with respect to the LDPC-based one in terms of decoding error probability. We think that this is due to the nature of the considered signals. Particularly, if we compare them just after the ADC, we find that their discrepancy is on average around 20$\%$ seeming too much for LDPC-based scheme to work properly.  On the other side, it is worth noting that the computational complexity of the Hamming distance decoding might become significantly huge when increasing the number of symbols taken as the encoder input and/or the number of quantization bits per symbol considered. 


Finally, to compare the proposed approach with respect to the literature, we consider as input the same dataset used in \cite{seepers}, i.e., the MIT-BIH Arrhythmia Database. It contains 48 half-hour excerpts of two-channel ambulatory ECG recordings, obtained from 47 subjects and digitized at 360 samples per second per channel with 11-bit resolution over a 10 mV range. These new ECG signals are less noisy with respect to \cite{ecg}. This turns out in a better detection of R peaks and, therefore, in less discrepancies between Alice and Bob's IPIs. Consequently, with this new dataset the $r$ significantly increases (up to $\frac{18}{72}$) while keeping a very small decoding error probability and finding the same key generation rate as in \cite{seepers}.

\section{Conclusions}
\label{sec:conc}

In  this  paper,  we  targeted  a  WBAN  collecting  ECG  signals from pairs of legitimate sensor nodes located on the individual’s chest, and we proposed  a  new  information  theoretic  key  agreement  protocol that exploits the inherent randomness of the ECG signals, independently acquired by the two nodes. The protocol is based on the capability of the sensors to correct the discrepancies between the IPIs sequence extracted by the ECG signals, in order to generate, independently, the same secret key.
We found that proper pre-processing is needed, before information reconciliation and privacy amplification steps. Specifically, the design of a quantizer that maximizes the source's entropy is necessary to limit the effectiveness of the attacker's guessing. 
Moreover, we observed that the complexity of the encoder (measured as the size of the parity check matrix) is strictly related to the key generation rate.
Finally, contrary to the expectation of needing a separate quantizer parity check matrix to fit the joint statistics of each ECG sensor pair, the similarity between them was such that one only quantizer, and parity check matrix, provided good performance for all ECG sensors pairs.

\end{document}